\documentclass[10pt]{article}
\usepackage{eurosym}
\usepackage{graphicx}
\usepackage{latexsym,graphicx}
\usepackage{float}
\usepackage{amssymb,amsmath,enumitem}
\usepackage{amsfonts}
\usepackage{amssymb}
\usepackage{epstopdf}
\usepackage{color}
\usepackage{cite}
\usepackage{hyperref}
\DeclareGraphicsExtensions{.eps}
\usepackage[left=2.5cm,top=2.5cm,right=2.5cm,bottom=2.5cm]{geometry}

\catcode `\@=11
\catcode `\@=12
\def\be{\begin{equation}}
	\def\ee{\end{equation}}
\def\beg{\begin{align}}
	\def\eeg{\end{align}}
\def\bea{\begin{eqnarray}}
	\def\eea{\end{eqnarray}}

\def\e{\mathrm{e}}

\begin{document}
	\begin{center}
		\large{\bf{Quintom-like transit universe models in Metric-affine $f(R,T,Q,T_m)$ gravity}} \\
		\vspace{5mm}
		\normalsize{ Dinesh Chandra Maurya$^{1}$, Harjit Kumar$^{2}$}\\
		\vspace{5mm}
		\normalsize{$^{1}$Centre for Cosmology, Astrophysics and Space Science, GLA University, Mathura-281 406,
			Uttar Pradesh, India.}\\
		\vspace{2mm}
		\normalsize{$^{2}$P.G. Department of Mathematics, S U College, Hilsa(Nalanda), Bihar, Patliputra University, Patna, Bihar.}\\
		\vspace{2mm}
		$^{1}$E-mail:dcmaurya563@gmail.com \\
		$^{2}$E-mail:harjeetkumar01@gmail.com \\
	\end{center}
	\vspace{5mm}
	\begin{abstract}
		The current transit universe model is a precise solution to the equations of a new type of gravity theory called metric-affine $f(R,T,Q,T_m)$ gravity proposed in [Herko et al. \textit{Phys. Dark Univ.} \textbf{34} (2021) 100886]. This theory is the maximal extension of the most successful theory, ``General Relativity," by including the scalars, Ricci curvature $R$, torsion $T$, nonmetricity $Q$, and trace $T_{m}$ of the matter-energy-momentum tensor using a generalized connection called the ``metric-affine" connection. We obtain the modified field equations for a linear form of the $f(R,T,Q,T_m)$ function and for a flat, homogeneous, and isotropic FLRW spacetime universe. We find a hyperbolic solution and determine the constrained values of the model parameters using the latest observational data. We examine how certain cosmological factors, like the deceleration parameter $q(z)$, effective equation of state parameter $\omega_{\rm eff}$, and dark energy equation of state parameter $\omega_{\rm de}$, vary over time to explain the properties of the observable universe. We perform the $Om$ diagnostic test for the model, and it represents the phantom scenarios of the model. The behavior of the dark energy EoS parameter $\omega_{\rm de}$ reveals the quintom-A-type universe characteristics.
	\end{abstract}
	\smallskip
	\vspace{5mm}
	\vspace{1cm}
	
	\section{Introduction}
		
	The rapid expansion of the universe during both early and late epochs \cite{ref1,ref2,ref3,ref4,ref5,ref6,ref7} poses challenges for explanation within the framework of general relativity (GR). This disparity between theory and observations has led to the development of various theories beyond GR, collectively known as ``Modified Gravity" \cite{ref8}. The exploration of a viable alternative has proven advantageous and informative for our comprehension of gravity. Modified gravity theories encompass metric formulations. Various gravitational theories include $f(R)$ theories, Metric-Affine (Palatini) $f(R)$ gravity \cite{ref9,ref10,ref11,sd1,sd2}, teleparallel $f(T)$ gravities \cite{ref12,ref13, jlsaid}, symmetric teleparallel $f(Q)$ \cite{ref14,ref15,ref16,ref17}, and scalar-tensor theories \cite{ref18,ref19} among others. The selection of alterations is mostly based on personal preference.  We believe that the most attractive options are those that add to the basic geometry of spacetime by providing a wider connection than the traditional Levi-Civita connection. In the absence of a priori restrictions on the connection, the space is typically not Riemannian \cite{ref20} and will exhibit both torsion and non-metricity. It is considered an additional fundamental field that overlays the metric. Determining the affine relationship facilitates the computation of the final geometric quantities. Metric-affine gravity theories have been formulated on this non-Riemannian manifold \cite{ref21}. Recently, \cite{ref22} examined $f(R)$ gravity theories utilizing symmetric, torsion-free connections. These are known as Palatini $f(R)$ theories of gravity. Reference \cite{ref23} examines the dynamics of metric-affine gravity theories, whereas reference \cite{ref24} analyzes the dynamics of generalized Palatini gravity theories. The metric-affine variational concepts in GR are examined in \cite{ref25}. Recently, the role of non-metricity in metric-affine gravity theories was investigated in \cite{ref26}.\\
	
	The Metric-Affine technique has become increasingly prominent in recent years, particularly in relation to its applications in cosmology \cite{ref28,ref29,ref30,ref31,ref32,ref33,ref34,ref35,ref36,ref37,ref38,ref39}. This attention may be due to a basic geometrical understanding of the other implications that operate in this context, unlike GR.  The observed alterations are entirely due to non-metricity and spacetime torsion.  Furthermore, geometric interpretations are encouraged by matter with intrinsic structure \cite{ref30} and \cite{ref40,ref41,ref42,ref43}. The connection between generalized geometry and inner structure offers a further benefit to the MAG approach. The affinely connected metric theories, especially the Riemann-Cartan subclass \cite{ref44}, justify the development of cosmological models utilizing a particular, albeit non-unique, connection. The simultaneous inclusion of non-zero curvature and torsion enhances the degrees of freedom necessary for gravitational alterations \cite{ref45}. The evolution of the early and late universes can be explained by metric-affine gravity \cite{ref46,ref47,ref48,ref49,ref50}.  This paradigm was used in a recent cosmology study \cite{ref46} to examine the temporal changes of observable quantities including density parameters and the effective dark energy equation-of-state parameter.  Using the mini-super-space technique, the study analyzed cosmic behavior, highlighting the influence of the connection, and introducing the theory as a distortion of GR and its teleparallel counterpart. Reference \cite{ref51} examined the observational constraints associated with Metric-Affine $F(R,T)$-gravity. Metric-affine gravity theories and their applications are detailed in \cite{source,ref52,ref53,ref55,ref56,ref57,ref58,dr1,dr2,dr3,dr4,dr5,dr6}.\\
	
	Motivated by the foregoing talks, we create some FLRW cosmological models in a generalized metric-affine matter-geometry coupling theory that was recently proposed by Harko et al. \cite{source}. We studied transit phase cosmological models in metric-affine $F(R,T)$ gravity with observational constraints \cite{ref60}. In \cite{ref61,ref62}, we investigated certain exact cosmological models in the metric-affine $F(R,T)$ gravity, while a dark energy model in metric-affine $f(R,Q)$ gravity is discussed in \cite{ref63}. In this study, we are considering different FLRW cosmological models and their attributes using the Metric-Affine $f(R,T,Q,T_m)$ gravity theory.\\
	
	The present paper is organized as follows:Section 2 contains some basic concepts and definitions used in the formulation of metric-affine gravity theory, and Section 3 contains a brief review of Metric-affine $f(R,T,Q,T_m)$ gravity theory. The field equations are given in section 4, while some cosmological exact solutions and observational analysis are presented in section 5. The result and discussions are shown in section 6. Finally, the conclusions are given in section 7.

	\section{Preliminary concepts}
	
	The current theory of gravity is formulated within a metric-affine manifold $\left(M, g_{ij}, \Gamma^{\rho}_{\,\,ij}\right)$. We define the covariant derivative of a vector $u^{i}$ as	
	\begin{equation}\label{p1}
		\nabla_{i}u^{\rho}=\partial_{i}u^{\rho}+\Gamma^{\rho}_{\;ji}u^{j}
	\end{equation}
	We also define the (Cartan)	torsion  tensor by \cite{ref53}
	\begin{equation}\label{p2}
		S_{ij}^{\;\;\;\;\rho}:=\Gamma^{\rho}_{\;\;\;[ij]}
	\end{equation}
	and the non-metricity tensor as
	\begin{equation}\label{p3}
		Q_{\rho ij} = \nabla_{\rho} g_{ij}
	\end{equation}
	Contracting these with the metric tensor we obtain the associated torsion and non-metricity vectors
	\begin{equation}\label{p4}
		S_{i}:=S_{ij}^{\;\;\;\;j}
	\end{equation}
	\begin{equation}\label{p5}
		Q_{i}:=Q_{i\alpha\beta}g^{\alpha\beta}\;\;,\;\; \tilde{Q}_{i}:=Q_{\alpha\beta i}g^{\alpha\beta}
	\end{equation}
	respectively. Consequently, we present the associated general metric-affine connection $\Gamma^{\rho}_{\,\,ij}$ as \cite{source}
	\begin{equation}\label{2}
		\Gamma^{\rho}_{\,\,ij}=\breve{\Gamma}_{\,\, ij}^{\rho}+K^{\rho}_{\,\,ij}+L^{\rho}_{\,\,ij}\,.
	\end{equation}
	In the context of Riemann geometry, the Levi--Civita connection $\breve{\Gamma}_{\,\, ij}^{\rho}$ is associated, while the contorsion tensor $K^{\rho}_{\,\,ij}$ relates to torsional geometry, and the disformation tensor $L^{\rho}_{\,\,ij}$ pertains to nonmetricity geometry.\\
	The definition of these tensors are as follows:
	\begin{align}
		\breve{\Gamma}^\alpha_{~\beta\gamma}=& \frac{1}{2} g^{\alpha\rho} \left( \partial _\gamma g_{\beta\rho} + \partial _\beta g_{\gamma\rho} - \partial _\rho g_{\beta\gamma} \right), \\
		K_{ij\alpha}=&\frac{1}{2}\bigl(T_{ij\alpha}+T_{j\alpha i}-T_{\alpha ij}\bigr),\\
		S_\alpha^{~ij}=&\frac12\bigl(K^{ij}_{~~~\alpha}+\delta^i_{\alpha} T^{\beta j}_{~~~\beta}-\delta^j_\alpha T^{\beta i}_{~~\beta}\bigr),\\
		L_{\alpha ij}=&\frac{1}{2}\bigl(Q_{\alpha ij}-Q_{ij\alpha}-Q_{ji\alpha}\bigr).
	\end{align}
	The non-metricity and torsion tensors have been added as
	\begin{equation}\label{3}
		\quad Q_{\rho ij} = \nabla_{\rho} g_{ij}, \,\,\,\, T_{\,\,\,\, ij}^{\alpha}=2 \Gamma_{\,\,\,\, [ij]}^{\alpha}.
	\end{equation}
	In a metric-affine spacetime, the introduction of the three scalars occurs as follows
	\begin{align}
		R=&g^{ij}R_{ij},\label{4}\\
		T=&{S_\rho}^{ij}\,{T^\rho}_{ij},\label{5}\\
		Q=& L^\alpha_{~\beta\alpha}L^{\beta i}_{~~~i}-L^\alpha_{~\beta i}L^{\beta i}_{~~~\alpha}.\label{6}
	\end{align}
	In this context, $R$ represents the curvature scalar associated with the Weyl-Cartan geometry, while $T$ denotes the torsion scalar, and $Q$ indicates the nonmetricity scalar.  The Ricci tensor is derived from the affine connection based on its conventional definition.
	\begin{align}\label{11}
		R_{j\beta}&=\partial_i\Gamma^i_{~j\beta}-\partial_j\Gamma^i_{~i\beta}+\Gamma^i_{~i\alpha}\Gamma^\alpha_{~j\beta}-\Gamma^i_{~j\alpha}\Gamma^\alpha_{~i\beta}.
	\end{align}	
	Then by using decomposition \eqref{2} we obtain the post Riemannian expansion for the Ricci scalar \cite{ref53}
	\begin{equation}
		R=\tilde{R}+T+Q +2 Q_{\alpha ij}S^{\alpha ij}+2 S_{i}(\tilde{Q}^{i}-Q^{i}) +\tilde{\nabla}_{i}(\tilde{Q}^{i}-Q^{i}-4S^{i})
	\end{equation}
	where $\tilde{R}$ is the Riemannian Ricci tensor (i.e. computed with respect to the Levi-Civita connection) and we have also defined the torsion and non-metricity scalars as
	\begin{equation}
		T:= S_{ij\alpha}S^{ij\alpha}-2S_{ij\alpha}S^{\alpha ij}-4S_{i}S^{i} \label{Tsc}
	\end{equation}
	and
	\begin{equation}
		Q:= \frac{1}{4}Q_{\alpha ij}Q^{\alpha ij}-\frac{1}{2}Q_{\alpha ij}Q^{ij\alpha}    -\frac{1}{4}Q_{i}Q^{i}+\frac{1}{2}Q_{i}\tilde{Q}^{i} \label{Qsc}
	\end{equation}
	respectively.
	
	\section{Metric-affine $f(R,T,Q,T_m)$ gravity}
	
	We analyze the following action for the metric-affine gravity theory $f(R,T,Q,T_m)$ \cite{source}.
	\begin{equation}\label{1}
		S=\int\sqrt{-g}d^{4}x \left[f\left(R,T,Q,T_m\right)+L_{m}\right],
	\end{equation}
	where $f$ represents a function dependent on the Ricci curvature scalar $R$, the torsion scalar $T$, the nonmetricity scalar $Q$, and the trace $T_{m}$ of the energy-momentum tensor (EMT) of the matter $T_{ij}$ derived from the matter Lagrangian $L_{m}$.  The gravitational theory formulated from the action \eqref{1} seeks to unify the various theories represented by $f(R)$ \cite{Bu1}, $f(T)$ \cite{TE5}, $f(Q)$ \cite{Q2}, $f\left(R,T_m\right)$ \cite{fT1}, $f\left(T,T_m\right)$ \cite{fTT}, and $f\left(Q,T_m\right)$ \cite{fQC1, fQC2}.\\
	
	By varying the action Eq.~(\ref{1}) with respect to both the connection and the metric, we derive the subsequent two field equations. \cite{M1a}
	\begin{align}\label{12}
		f_R&\Big[g_{i[j}Q_{\alpha]\beta}^{~~~~\beta}+T_{i\alpha j}+Q_{\alpha ij}-g_{\alpha i}Q_{\beta j}^{~~~\beta}\Big]+f_Q\Big[2g_{i(j}L^\beta_{~\alpha)\beta}+g_{\alpha j}(L_{i\beta}^{~~~\beta}-L^\beta_{~i\beta})-2L_{i\alpha j}\Big]\nonumber\\
		&+f_T\Big[T_{\alpha ij}+T_{i\alpha j}-T_{j\alpha i}+g_{\alpha[j}T^\beta_{~~i]\beta}\Big]+2g_{i[\alpha}\nabla_{j]}f_R-\frac12H_{\alpha ij}=0,
	\end{align}
	and
	\begin{align}\label{13}
		f_R R_{ij}&-\frac12f\,g_{ij}+f_{T_m}(g_{ij}L_m-T_{ij})-\frac12T_{ij}-\frac12\nabla_\alpha\Big(A^\alpha_{~(ij)}-A_{(i~~j)}^{~~\,\alpha}+A_{(ij)}^{~~~~\alpha}\Big)\nonumber\\
		&+\frac14f_T\Big[2T_{\alpha j\beta}T^{\alpha~~\beta}_{~~i}+2T^{\alpha~~\beta}_{~~i}T_{\beta j}^{~~~\alpha}-4T^\alpha_{~~i\alpha}T^\beta_{~~j\beta}-T_i^{~~\alpha\beta}T_{j\alpha\beta}\Big]\nonumber\\
		&+\frac14f_Q\Big[3Q_{\alpha~~\beta}^{~~\beta}Q^\alpha_{~ij}-g_{ij}Q_{\alpha\beta}^{~~~\beta}Q^{\alpha\gamma}_{~~~\gamma}-2Q_{\alpha j\beta}Q^{\alpha~~\beta}_{~\,i}+2Q^{\alpha~~\beta}_{~~i}Q_{\beta j\alpha}-2Q^\alpha_{~i\alpha}Q^\beta_{~j\beta}\nonumber\\
		&+Q_{\beta(i}^{~~~~\beta}Q_{j)\alpha}^{~~~~\alpha}-2Q_{\alpha~~\beta}^{~~\beta}Q_{(ij)}^{~~~~\alpha}+Q_i^{~~\alpha\beta}Q_{j\alpha\beta}\Big]=0,
	\end{align}
	correspondingly, where the energy-momentum tensor and the hypermomentum tensor are defined by
	\begin{eqnarray}\label{14}
		{ T}_{ij}=-\frac{2}{\sqrt{-g}}\frac{\delta (\sqrt{-g}L_{m})}{\delta g^{ij}}, \quad H_{\lambda}^{\, \, \, ij}=-\frac{2}{\sqrt{-g}}\frac{\delta (\sqrt{-g}L_{m})}{\delta \Gamma^{\lambda}_{\,\,\, ij}}.
	\end{eqnarray}
	Furthermore, we have established
		\begin{align}\label{15}
			A_{i\alpha j}=f_Q\big(g_{ij}L^\beta_{~\alpha\beta}+g_{\alpha j}L_{i\beta}^{~~\beta}-L_{i\alpha j}-L_{j\alpha i}\big).
	\end{align}
	All curvature terms and derivatives are derived from the affine connection $\Gamma^\alpha_{~ij}$.\\
	
	We will now present the Weyl vector, beginning with a straightforward formulation of the non-metricity tensor, which we will define as follows.
	\begin{align}\label{16}
		Q_{ij\alpha}=w_i g_{j\alpha}.
	\end{align}
	In this context, $w_{i}$ represents the Weyl vector.  The gravitational field equations of the $f\left(R,T,Q,T_m\right)$ theory, in this instance, reduce to
	\begin{align}\label{17}
		f_T&\Big[T_{\alpha ij}-g_{\alpha[i}T^\beta_{~~j]\beta}+T_{i\alpha j}-T_{j\alpha i}\Big]\nonumber\\&+f_Q\Big[2g_{\alpha(j} w_{i)}-2g_{i(\alpha}w_{j)}\Big]\nonumber\\&+f_R\Big[T_{i\alpha j}-2g_{i[\alpha}w_{j]}\Big]+2g_{i[j}\nabla_{\alpha]}f_R-\frac12H_{\alpha ij}=0,
	\end{align}
	and
	\begin{align}\label{18}
		f_R R_{ij}&-\frac12f\,g_{ij}+f_{T_m}(g_{ij}L_m-T_{ij})-\frac12T_{ij}-g_{ij}w^\alpha\nabla_\alpha f_Q+w_{(j}\nabla_{i)}f_Q\nonumber\\&
		+\frac12f_Q\Big[\nabla_{(i}w_{j)}-5g_{ij}w^\alpha w_\alpha-w_i w_j-g_{ij}\nabla_\alpha w^\alpha-g_{\alpha(i}\nabla_{j)}w^\alpha\Big]
		\nonumber\\&+\frac14f_T\Big[2\,T_{\alpha j\beta}T^{\alpha~~\beta}_{~~i}+2\,T^{\alpha~~\beta}_{~~i}T_{\beta j}^{~~~\alpha}-4T^\alpha_{~~i\alpha}T^\beta_{~~j\beta}-T_i^{~~\alpha\beta}T_{j\alpha\beta}\Big]=0,
	\end{align}
	in that order.  At this juncture, it is important to observe that in the equations presented above, all curvature terms and their derivatives are derived from the affine connection $\Gamma^\alpha_{~ij}$.\\
	
	The torsion tensor $T_{ij\alpha}$ can be expressed in a decomposed form as
	\begin{align}\label{19}
		T_{ij\rho}=\frac23(t_{ij\rho}-t_{i\rho j})+(A_j g_{i\rho}-A_\rho g_{ij})+\epsilon_{ij\rho\sigma}B^\sigma,
	\end{align}
	where $t_{ij\alpha}$ represents a tensor characterized by the following properties,
	\begin{align}\label{20}
		t_{ij\rho}+t_{j\rho i}+t_{\rho ij}=0,\quad t^i_{~\,ij}=0=t^i_{~\,ji},
	\end{align}
	and $3A_i=T^\alpha_{~i\alpha}$, where $B_i$ represents two arbitrary vectors.  This study will operate under the assumption that solely the $A_i$ vector is non-zero.  The torsion tensor can therefore be expressed as
	\begin{align}\label{21}
		T_{ij\alpha}=A_j g_{i\alpha}-A_\alpha g_{ij}.
	\end{align}
	
	The equations governing the field exhibit simplification in this scenario as
	\begin{align}\label{22}
		8&f_T g_{\alpha[i}A_{j]}+2f_Q\Big[g_{\alpha(j}w_{i)}-g_{i(j}w_{\alpha)}\Big]\nonumber\\&+2f_R\Big[g_{i[j}A_{\alpha]}-g_{i[\alpha}w_{j]}\Big]+2g_{i[\alpha}\nabla_{j]}f_R=0,
	\end{align}
	and
	\begin{align}\label{23}
		f_R \breve{R}_{ij}&-\frac12f\,g_{ij}+f_{T_m}(g_{ij}L_m-T_{ij})-\frac12T_{ij}-g_{ij}w^\alpha\breve{\nabla}_\alpha f_Q+w_{(j}\breve{\nabla}_{i)}f_Q-f_T\,A_i A_j\nonumber\\&
		+\frac12f_Q\Big[3A^\alpha w_\alpha g_{ij}-2g_{ij}w_\alpha w^\alpha-3A_{(i}w_{j)}+2w_i w_j-2g_{ij}\breve{\nabla}_\alpha w^\alpha+2\breve{\nabla}_{(i}w_{i)}\Big]\nonumber\\&
		+\frac12 f_R\Big[2(A_\alpha A^\alpha g_{ij}-A_i A_j)+A^\alpha w_\alpha g_{ij}+2A_{(i}w_{j)}+w_i w_j-w^\alpha w_\alpha g_{ij}\nonumber\\&-g_{ij}\breve{\nabla}_\alpha(2A^\alpha-w^\alpha)+\breve{\nabla}_{(i}A_{j)}+\breve{\nabla}_{(i}w_{j)}\Big]=0,
	\end{align}
	in that order. The equations provided include all curvature terms and derivatives obtained from the Levi-Civita connection $\breve{\Gamma}^\alpha_{~ij}$. On the other hand, we have proposed that the energy-momentum tensor of the matter field is derived exclusively from the metric tensor and is not influenced by the affine connection.\\
	
	For future reference, we can note that the torsion, non-metricity, and curvature scalars can be expressed as
	\begin{align}\label{24}
		T&=-6A^2,\quad Q=-\frac32 w^2,\nonumber\\
		R&=\breve{R}+3A^2+3A_\alpha w^\alpha-\frac32w^2-3\breve{\nabla}_i A^i+3\breve{\nabla}_i w^i,
	\end{align}
	with $A^2\equiv A_\alpha A^\alpha$ and $w^2\equiv w_\alpha w^\alpha$.
	
\section{Field equations}

The Friedmann-Lema\^{\i}tre-Robertson-Walker (FLRW) metric is examined in its conformal coordinates representation, which is expressed as follows
\begin{equation}\label{25}
	ds^2=a^2(t)\left(-dt^2+d\vec{x}^2\right),
\end{equation}
where $a=a(t)$ represents the conformal scale factor.  The Hubble parameter is introduced, defined as $H=\dot{a}/a$.  It is posited that the matter sector of the Universe can be characterized by a perfect fluid, with the equations of motion derivable from the Lagrangian $L_m=-\rho$, and which possesses the energy-momentum tensor.
\begin{align}\label{26}
	T^\mu_{~\nu}=\textmd{diag}(-\rho,p,p,p).
\end{align}
In this context, $\rho$ represents the energy density, while $p$ denotes the thermodynamic pressure.\\

Given the homogeneity and isotropy of space-time, the expressions for the Weyl and torsion vectors can be formulated as
\begin{align}\label{27}
	A_{\mu}=(aA_0,\vec{0}),\nonumber\\
	w_\mu=(aw_0,\vec{0}),
\end{align}
where $A_0=A_0(t)$ and $w_0=w_0(t)$ represent two functions that vary with time.  The equation of motion pertaining to the affine connection is presented here.  Equation \eqref{22} consists of two independent components, which correspond to the Weyl vector and the torsion vector. These can be simplified as follows:
\begin{align}\label{28}
	&f_R(A_0+w_0)-f_Q w_0=\frac{1}{a}\dot{f}_R,\nonumber\\
	&4f_TA_0-f_Qw_0=0.
\end{align}

The generalized Friedman and Raychaudhuri equations can be derived as follows \cite{M1a}
\begin{equation}\label{29}
	a^2\Big[f+2A_0^2f_T-3A_0w_0f_R\Big]+a\Big[\dot{w}_0+Hw_0\Big]f_{Q}+a\Big[\dot{A}_0-\dot{w}_0+H(A_0-w_0)\Big]f_{R}-6(\dot{H}+H^{2})f_R=a^2\rho
\end{equation}
and
\begin{eqnarray}\label{30}
	a^2f-2(\dot{H}+3H^2)f_R+2a^2(\rho+p)f_{T_m}+2aw_0\dot{f}_Q+a\Big[2\dot{w}_0+w_0(2H+2aw_0-3aA_0)\Big]f_Q\nonumber\\
	-a\Big[\dot{w}_0-2\dot{A}_0+a(2A_0-w_0)(A_0+w_0)-(2A_0-w_0)H\Big]f_R=-a^2p,
\end{eqnarray}
respectively.

\section{Cosmological solutions}

In this study, to solve the above complicated field equations, we consider the linear form of arbitrary function $f(R,T,Q,T_m)$ as \cite{source}
\begin{equation}\label{31}
	f(R,T,Q,T_m)=\frac{1}{2}R+\alpha\,T+\beta\,Q+\gamma\,T_{m},
\end{equation}
where $\alpha$, $\beta$ and $\gamma$ are coupling arbitrary constants. From Eq.\,\eqref{31}, we get the partial derivatives as
\begin{equation}\label{32}
	f_{R}=\frac{1}{2},~~~~~f_{T}=\alpha,~~~~~~f_{Q}=\beta,~~~~~~f_{T_{m}}=\gamma.
\end{equation}
Using Eq.\,\eqref{32} in \eqref{28}, we obtain the following relations
\begin{align}\label{33}
	A_{0}-(2\beta-1)w_{0}=0,\nonumber\\
	4\alpha\,A_{0}-\beta\,w_{0}=0.
\end{align}
Solving Eq.\,\eqref{33}, we obtain the solution either
	\begin{equation}\label{34}
		A_{0}=0, w_{0}=0,
	\end{equation}
	or
	\begin{equation}\label{35}
		A_{0}=(2\beta-1)k,~~~~w_{0}=k,
	\end{equation}
with condition $\alpha=\frac{\beta}{2\beta-1}$.\\
In the first case, the solution \eqref{34} reveals that both torsion and nonmetricity vanishes together, and hence, we get $f(R,T,Q,T_m)=\frac{1}{2}R+\gamma\,T_{m}$ which is a linear form of $f(R,T_{m})$ theory and it is widely investigated in past literature. One can obtain the GR case for $\alpha=0, \beta=0, \gamma=0$ with $A_{0}=0$, and $w_{0}=0$. In \cite{source}, the authors have considered the first case without investigating observational constraints, and we considered a general solution with observational constraints, which presents a model of dark energy without adding a cosmological constant or constant in the Lagrangian function.\\
Hence, in this study, we consider the solution \eqref{35} and in this case, the function $f(R,T,Q,T_m)=\frac{1}{2}R+\frac{\beta}{2\beta-1}\,T+\beta\,Q+\gamma\,T_{m}$. Using this form of solution, Eqs.\,\eqref{29} and \eqref{30} reduce to
	\begin{equation}\label{36}
		3H^{2}-(\beta-2)kH+\frac{k^{2}}{2}(4\beta^{2}-7\beta+1)=(1+\gamma)\rho-3\gamma\,p
	\end{equation}
	and
	\begin{equation}\label{37}
		2\dot{H}+3H^{2}-\frac{3}{2}(2\beta-3)kH-\frac{k^{2}}{2}(4\beta^{2}-3\beta+2)=-(1+5\gamma)p-\gamma\,\rho,
	\end{equation}
	respectively.\\
We choose dust fluid background source $p=0$, then from Eqs.\,\eqref{36} and \eqref{37}, we get
\begin{equation}\label{38}
	\dot{H}+k_{1}\,H^{2}-k_{2}\,H-k_{3}=0,
\end{equation}
with $k_{1}=\frac{3(1+2\gamma)}{2(1+\gamma)}$, $k_{2}=\frac{k[2(3+4\gamma)\beta-(9+13\gamma)]}{4(1+\gamma)}$ and $k_{3}=\frac{k^{2}[4\beta^{2}-3\beta+2+\gamma(8\beta^{2}-10\beta+3)]}{2(1+\gamma)}$.\\
Solving Eq.\,\eqref{38}, we get the following Hubble function:
\begin{equation}\label{39}
	H(t)=\frac{k_{2}}{2k_{3}}+\frac{k_{4}}{k_{3}}\coth[k_{4}(t-c_{0})],
\end{equation}
where $k_{4}=\frac{1}{2}\sqrt{k_{2}^{2}+4k_{1}k_{3}}$ and $c_{0}$ is an integrating constant.\\
Now, integrating Eq.\,\eqref{39}, we get the scale factor $a(t)$ as
\begin{equation}\label{40}
	a(t)=\e^{k_{5}(t-c_{1})}\left(\sinh[k_{4}(t-c_{0})]\right)^{\frac{1}{k_{3}}},
\end{equation}
where $k_{5}=\frac{k_{2}}{2k_{3}}$ and $c_{1}$ is an integrating constant.\\

Now, for $k_{2}=0$, we get the hyperbolic solution
\begin{equation}\label{41}
	a(t)=\left(\sinh[\sqrt{k_{1}k_{3}}(t-c_{0})]\right)^{\frac{1}{k_{3}}},
\end{equation}
and for $\gamma=-1$, we get exponential solution
\begin{equation}\label{42}
	a(t)=\e^{k_{5}(t-c_{1})}.
\end{equation}
Here, we have considered hyperbolic solution \eqref{41}, and hence, we have obtained the Hubble parameter as
\begin{equation}
	H(t)=\sqrt{\frac{k_{1}}{k_{3}}}\coth[\sqrt{k_{1}k_{3}}(t-c_{0})],
\end{equation}
with $\beta=\frac{9+13\gamma}{6+8\gamma}$.\\
Using the relationship $a^{-1}=1+z$, we get the Hubble function in terms of $z$ as
\begin{equation}\label{hz}
	H(z)=\frac{H_{0}}{\sqrt{2}}\sqrt{1+(1+z)^{2k_{3}}},
\end{equation}
where $H_{0}=\sqrt{\frac{2k_{1}}{k_{3}}}$.\\
From Eqs.\,\eqref{36} and \eqref{37}, we obtain the energy density $\rho$ as
\begin{equation}
	\rho=3H_{0}^{2}-\frac{1}{2}k^{2}(4\beta+1)+(3-k_{3})H_{0}^{2}(1+z)^{2k_{3}}-\frac{1}{2\sqrt{2}}(8\beta-13)kH_{0}\sqrt{1+(1+z)^{2k_{3}}},
\end{equation}
and total matter energy density parameter $\Omega$ is obtained as
\begin{equation}
	\Omega=1-\frac{k_{3}(1+z)^{2k_{3}}}{3[1+(1+z)^{2k_{3}}]}-\frac{k(8\beta-13)}{6\sqrt{2}H_{0}\sqrt{1+(1+z)^{2k_{3}}}}-\frac{(4\beta+1)k^{2}}{6H_{0}^{2}[1+(1+z)^{2k_{3}}]}.
\end{equation}
Now, Eqs.\,\eqref{36} and \eqref{37} can be rewritten as
\begin{equation}
	3H^{2}=\rho_{\rm m}+\rho_{\rm de},
\end{equation}
and
\begin{equation}
	2\dot{H}+3H^{2}=-p_{\rm m}-p_{\rm de},
\end{equation}
where $\rho_{\rm m}=\rho$, $p_{\rm m}=p$, and dark energy density $\rho_{\rm de}$ and pressure $p_{\rm de}$ is defined as
\begin{equation}
	\rho_{\rm de}=\gamma\rho-(\beta-2)kH-\frac{k^{2}}{2}(4\beta^{2}-7\beta+1),
\end{equation}
and 
\begin{equation}
	p_{\rm de}=\gamma\rho-\frac{3}{2}(2\beta-3)kH-\frac{k^{2}}{2}(4\beta^{2}-3\beta+2).
\end{equation}
We have derived the dark energy density $\rho_{\rm de}$ and pressure $p_{\rm de}$ as
\begin{equation}
\rho_{\rm de}=k_{3}H_{0}^{2}(1+z)^{2k_{3}}+\frac{1}{2\sqrt{2}}kH_{0}(8\beta-13)\sqrt{1+(1+z)^{2k_{3}}}+\frac{1}{2}(4\beta+1),
\end{equation}
and
\begin{equation}
	p_{\rm de}=-3H_{0}^{2}-(3-k_{3})H_{0}^{2}(1+z)^{2k_{3}}.
\end{equation}
Now, the dark energy EoS parameter $\omega_{\rm de}$ is obtained as
\begin{equation}\label{wde}
\omega_{\rm de}=\frac{-3H_{0}^{2}-(3-k_{3})H_{0}^{2}(1+z)^{2k_{3}}}{k_{3}H_{0}^{2}(1+z)^{2k_{3}}+\frac{1}{2\sqrt{2}}kH_{0}(8\beta-13)\sqrt{1+(1+z)^{2k_{3}}}+\frac{1}{2}(4\beta+1)}.
\end{equation}

\subsection{Observational constraints}
In order to determine the optimal values for the model parameters $H_{0}$ and $\gamma$ using observational datasets, we conduct a Monte Carlo Markov Chain (MCMC) analysis on a dataset comprising $31$ cosmic chronometers (CC) and $1048$ from the Pantheon sample of SNe Ia. This analysis is facilitated by the emcee software, which is freely accessible at \cite{ref71}.

\subsection*{Hubble data}
In the exploration of the cosmic expansion history of the universe, the Hubble data points play a vital role. We utilize $31$ data points of $H(z)$ from the CC sample \cite{cc1,cc2} within the range of $0.07\le z \le 1.965$. The data points exhibit no correlation and can be analyzed using the following $\chi^{2}$ formula:
\begin{equation}\label{eq26}
	\chi_{CC}^{2}=\sum_{i=1}^{i=31}\frac{[H_{ob}(z_{i})-H_{th}(H_{0}, \gamma, z_{i})]^{2}}{\sigma_{H(z_{i})}^{2}}.
\end{equation}
where $H_{0}, \gamma$ are the cosmological parameters which we have to estimate, and $H_{ob}$, $H_{th}$ are the observational and theoretical values of $H(z)$ at $z=z_{i}$, respectively. The $\sigma_{H(z_{i})}$ denotes the standard deviations associated with observed values $H_{ob}$.\\

Figure 1 illustrates the 2D contour plots of model parameters $H_{0}$ and $\gamma$ at the $\sigma1$ and $\sigma2$ levels of confidence. The estimated constrained values of $H_{0}$ and $\gamma$, along with the minimum $\chi$ square value, are presented in Table \ref{T1}. The value of the Hubble constant has been measured as $H_{0}=65.3_{-2.5}^{+2.1}$ km/s/Mpc for the CC dataset, aligning with recent observations.
\begin{figure}[H]
	\centering
	\includegraphics[width=8cm,height=8cm,angle=0]{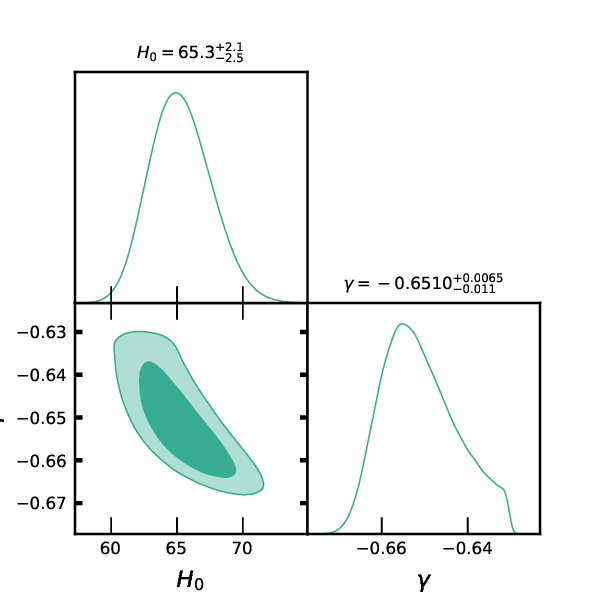}
	\caption{2D contour plots of $H_{0}, \gamma$ using MCMC analysis of $31$ CC dataset.}
\end{figure}

\subsection*{Pantheon data}

Type Ia supernova data is employed to quantify the expansion rate of the universe's cosmic evolution, expressed as apparent magnitude $m(z)$. Here, we use the $1048$ data points of apparent magnitude from Pantheon sample of SNe Ia as reported in \cite{ref79}. We examined the theoretical concept of apparent magnitude, as detailed in \cite{ref74,ref75,ref76,ref77, as1,ref78}.
\begin{equation}\label{eq27}
	m(z)=M+ 5~\log_{10}\left(\frac{D_{L}}{Mpc}\right)+25.
\end{equation}
where $M$ denotes the absolute magnitude, and the luminosity distance $D_{L}$ is measured in units of length and is defined as
\begin{equation}\label{eq28}
	D_{L}=c(1+z)\int^z_0\frac{dz'}{H(z')}.
\end{equation}
The Hubble-free luminosity distance \( d_{L} \) is defined as \( d_{L} \equiv \frac{H_{0}}{c}D_{L} \), representing a dimensionless quantity. Thus, the apparent magnitude $m(z)$ is defined as
\begin{equation}\label{eq29}
	m(z)=M+5\log_{10}\left(\frac{c/H_{0}}{Mpc}\right)+25+5\log_{10}{d_{L}},
\end{equation}
\begin{equation}\label{eq30}
	m(z)=\mathcal{M}+5\log_{10}{d_{L}}.
\end{equation}
Our analysis revealed a degeneracy between $H_{0}$ and $M$ in the previously reported equation, which maintains its invariance within the $\Lambda$CDM framework \cite{ref74,ref77}. We will integrate these degenerate parameters through a redefinition as outlined below:
\begin{equation}\label{eq31}
	\mathcal{M}\equiv M+5\log_{10}\left(\frac{c/H_{0}}{Mpc}\right)+25.
\end{equation}
In this context, $\mathcal{M}$ represents a dimensionless parameter, defined by the equation $\mathcal{M}=M-5\log_{10}(h)+42.39$, where $H_{0}=h\times100 \text{ km/s/Mpc}$. Within the framework of the $\Lambda$CDM model, the parameter $\mathcal{M}=23.809\pm0.011$ has been calibrated as outlined in reference \cite{ref74}. The parameter $\mathcal{M}$ shows differences among different cosmological theories (see \cite{ref74,ref80,ref81,ref82,ref83,ref84,ref85,ref86,ref87,ref88,ref89,ref90}). The examination of the Pantheon data is carried out utilizing the subsequent $\chi^{2}$ formula, as cited in \cite{ref74}:
\begin{equation}\label{eq32}
	\chi^{2}_{P}=V_{P}^{i}C_{ij}^{-1}V_{P}^{j}
\end{equation}
The formula \( V_{P}^{i} \) demonstrates the difference between the observed value \( m_{ob}(z_{i}) \) and the expected value \( m(\gamma, \mathcal{M}, z_{i}) \) as outlined in equation \eqref{eq30}.   The notation \( C_{ij}^{-1} \) represents the inverse of the covariance matrix derived from the Pantheon sample.\\
For the joint analysis of CC+Pantheon $1079$ datasets, we use the following $\chi$ square formula:
\begin{equation}\label{eq33}
	\chi_{CC+P}^{2}=\chi_{CC}^{2}+\chi^{2}_{P}
\end{equation}
\begin{table}[H]
	\centering
	\begin{tabular}{|c|c|c|c|}
		\hline
		
		Parameter        & Prior                & CC                         & CC$+$Pantheon\\
		\hline
		$H_{0}$          & $(50, 100)$          & $65.3_{-2.5}^{+2.1}$       & $69.0\pm1.8$\\
		$\gamma$         & $(-0.99, -0.63)$     & $-0.6510_{-0.011}^{+0.0065}$& $-0.6648_{-0.0027}^{+0.0022}$\\
		$\mathcal{M}$    & $(23, 24)$           & -                          & $23.8310\pm0.0090$\\
		$\chi^{2}$       & -                    & $14.9005$                  & $1047.7762$\\
		\hline
	\end{tabular}
	\caption{The MCMC estimates.}\label{T1}
\end{table}
Figure 2 presents the 2D contour plots illustrating the model parameters $H_{0}$, $\gamma$, and $\mathcal{M}$ at the confidence levels of $\sigma1$ and $\sigma2$.  The constrained estimates for $H_{0}$, $\gamma$, and $\mathcal{M}$, along with the corresponding minimum $\chi$ square value, are presented in Table \ref{T1}.  The value of the Hubble constant has been measured as $H_{0}=69.0\pm1.8$ km/s/Mpc for the CC+Pantheon dataset, aligning with recent observations.  Recently, in several investigations, cosmologists reported the value of the Hubble constant as $H_{0}=68.326_{-1.045}^{+1.005}$ km/s/Mpc in \cite{h1}, $H_{0}=68.95_{-0.95}^{+2.17} $ km/s/Mpc in \cite{h2}, $H_{0}=64.75 $ km/s/Mpc in \cite{h3}, $H_{0}=67.71\pm0.86 $ km/s/Mpc in \cite{h4}, $H_{0}=68_{-2.0}^{+2.3}$ km/s/Mpc in \cite{h5}.  Additional measurements of the Hubble constant yield a value of $H_{0}=69.8\pm1.3$ km/s/Mpc in \cite{h6}, $H_{0}=69.7\pm1.2$ km/s/Mpc in \cite{h7}, $H_{0}=68.3721\pm1.7205$ km/s/Mpc in \cite{h8}, $H_{0}=68.3721\pm1.65678$ km/s/Mpc in \cite{h9}, and $H_{0}=71.66123\pm0.33061$ in \cite{h10}.  Consequently, our estimated values for the Hubble constant fall within the range of $64< H_{0}<73$ based on the analysis of both datasets.
\begin{figure}[H]
	\centering
	\includegraphics[width=10cm,height=10cm,angle=0]{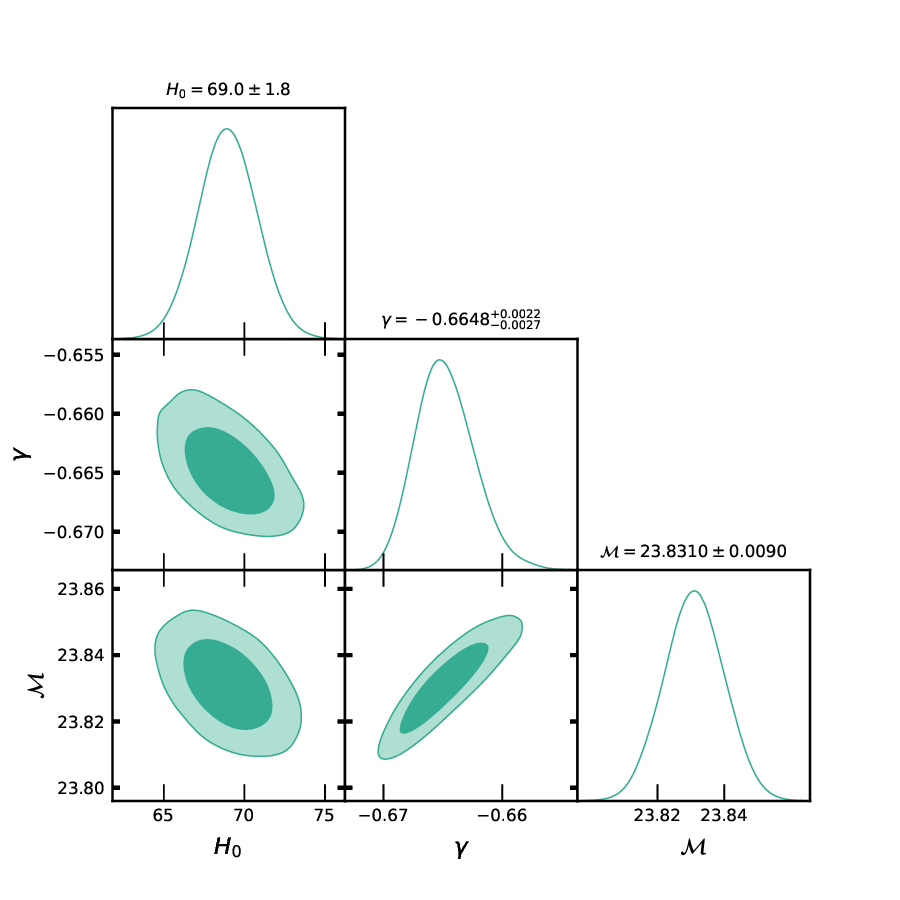}
	\caption{2D contours of $H_{0}$, $\gamma$ and $\mathcal{M}$ at $\sigma1$ and $\sigma2$ confidence level for CC+Pantheon datasets.}
\end{figure}

\section{Result discussions}
In this section, we will discuss the physical behavior of cosmological parameters and their implications in evolution of universe. The effective equation of state (EoS) is defined as $\omega_{\rm eff}=-1+\frac{2}{3}(1+z)\frac{H'}{H}$ and is derived for the model as
\begin{equation}\label{wef}
	\omega_{\rm eff}=-1+\frac{2}{3}\frac{k_{3}(1+z)^{2k_{3}}}{1+(1+z)^{2k_{3}}}
\end{equation}
where $k_{3}=\frac{k^{2}[4\beta^{2}-3\beta+2+\gamma(8\beta^{2}-10\beta+3)]}{2(1+\gamma)}$ and $\beta=\frac{9+13\gamma}{6+8\gamma}$.\\
We have measured the value of the Hubble constant $H_{0}=65.3_{-2.5}^{+2.1}$ km/s/Mpc and the model parameter $\gamma=-0.6510_{-0.011}^{+0.0065}$ for the CC dataset, while for the CC+Pantheon dataset, we measured $H_{0}=69.0\pm1.8$ km/s/Mpc and $\gamma=-0.6648_{-0.0027}^{+0.0022}$. We have used the value of the arbitrary constant $k=1$ in our analysis and discussion of the results. We have obtained the relationships between $\alpha$, $\beta$, and $\gamma$ as $\alpha=\frac{9+13\gamma}{12+18\gamma}$ and $\beta=\frac{9+13\gamma}{6+8\gamma}$. So, by using the value of $\gamma$, we can calculate the constrained values of the other model parameters $\alpha$ and $\beta$ as $\alpha=0.6780_{-0.1184}^{+0.0583},~0.5246_{-0.0637}^{+0.0277}$ and $\beta=1.9042_{-0.3466}^{+2.7862},~10.6428_{-2.1803}^{+0.3056}$, based on two sets of observational data, CC and CC+Pantheon.\\
\begin{figure}[H]
	\centering
	\includegraphics[width=10cm,height=8cm,angle=0]{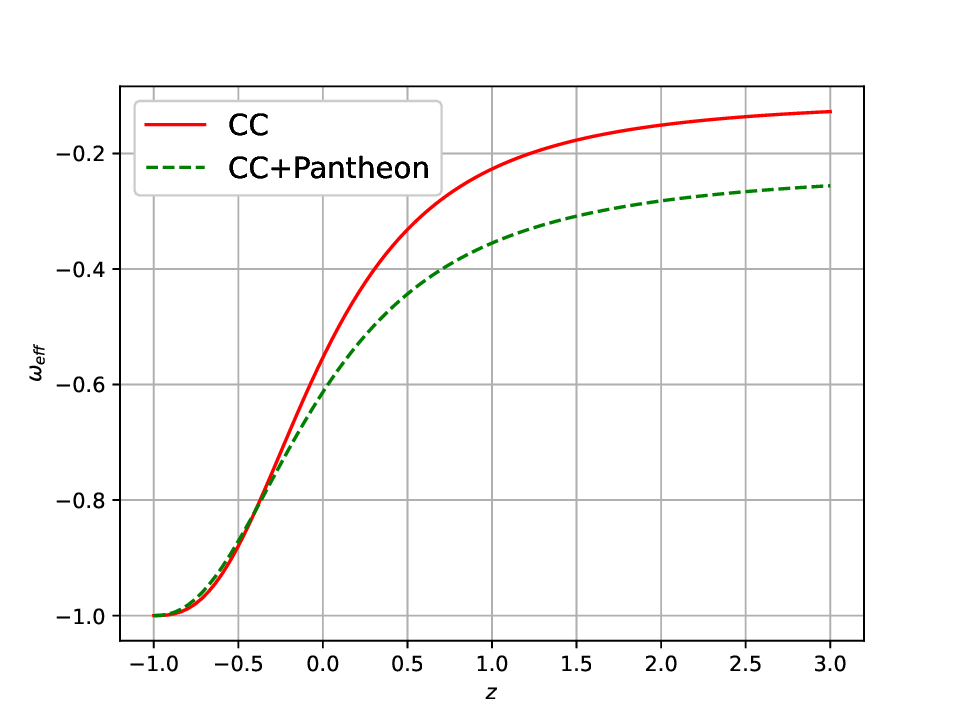}
	\caption{The variation of effective EoS parameter $\omega_{\rm eff}$ versus $z$.}
\end{figure}
\begin{figure}[H]
	\centering
	\includegraphics[width=10cm,height=8cm,angle=0]{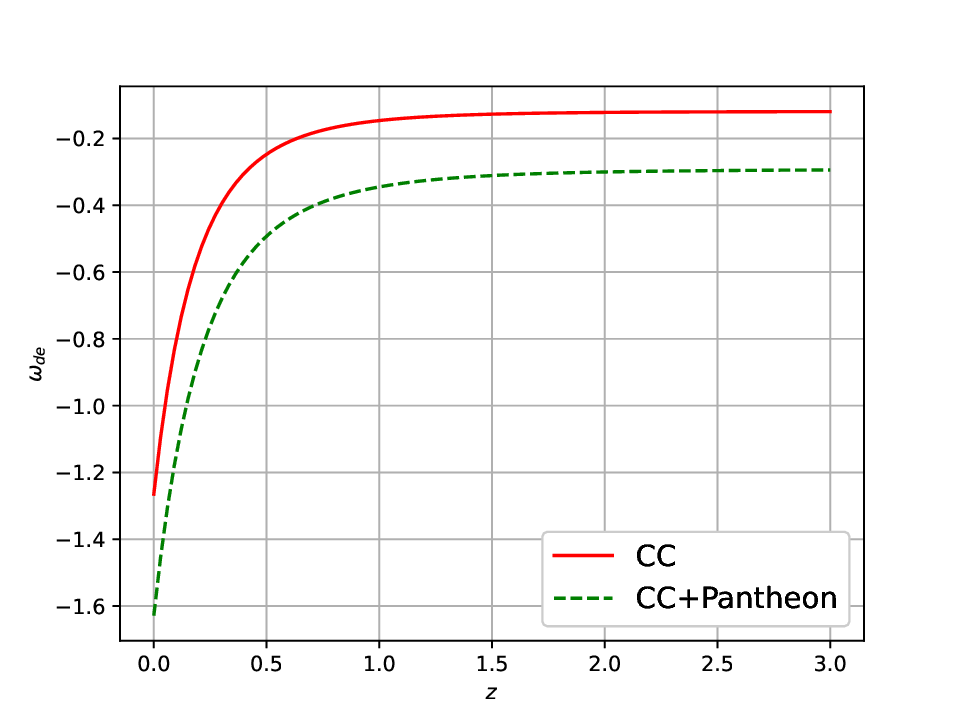}
	\caption{The variation of dark energy EoS parameter $\omega_{\rm de}$ versus $z$.}
\end{figure}
Figure 3 illustrates the plot of the effective EoS parameter $\omega_{\rm eff}$ using the above estimated values of model parameters. From Figure 3, we can see that the effective EoS parameter varies as $-1\le\omega_{\rm eff}<0$ over the redshift $z\in[-1,\infty)$. Also, one can see that as $z\to\infty$, then $\omega_{\rm eff}\to-1+\frac{2}{3}k_{3}$, which shows that early scenarios of the model depend upon the value of model parameters $\alpha$, $\beta$, and $\gamma$. We found that the current value of the effective EoS parameter is $\omega_{\rm eff}=-0.55\pm0.16$ using CC data, and when combining CC with Pantheon data, it is estimated to be $\omega_{\rm eff}=-0.62\pm0.11$, which is similar to the $\Lambda$CDM value of about $\omega_{\rm eff}\approx-0.70$ \cite{ref2,ref3}.\\

Equation \eqref{wde} represents the expression for the dark energy EoS parameter, and its variation over redshift $z$ is shown in Figure 4. From Figure 4, one can observe that the evolution of $\omega_{\rm de}$ over $z$ illustrates the quintom-type scenarios of the universe's evolution. Recently, the quintom-type scenarios are extensively discussed and categorized in \cite{quintom1,quintom2,quintom3,quintom4}. In \cite{quintom1}, it was shown that there are two main types of quintom models: quintom-A, where a normal field is more important in the early universe and a phantom field takes over in the late universe, leading to a change from above to below the phantom divide; and quintom-B, where the equation of state moves from below $\omega=-1$ to above $\omega=-1$. Thus, one can see that Figure 4 shows quintom-A-like scenarios of the model. The dark energy equation of state $\omega_{\rm de}$ is reported as $\omega_{\rm de}=-0.98\pm0.12$ in \cite{spergel2003}, while it is measured as $\omega_{\rm de}=-1.0\pm0.19$ and $\omega_{\rm de}=-1.10\pm0.14$ in \cite{komatsu2011}. We have measured the present value of this dark energy EoS parameter as $\omega_{\rm de}=-1.2648$ along CC data, while for CC+Pantheon datasets $\omega_{\rm de}=-1.6285$.\\

The deceleration parameter is defined as $q(z)=-1+(1+z)\frac{H'}{H}$ and is derived for the model as
\begin{equation}\label{dec}
	q(z)=-1+\frac{k_{3}(1+z)^{2k_{3}}}{1+(1+z)^{2k_{3}}}
\end{equation}
where $k_{3}=\frac{k^{2}[4\beta^{2}-3\beta+2+\gamma(8\beta^{2}-10\beta+3)]}{2(1+\gamma)}$ and $\beta=\frac{9+13\gamma}{6+8\gamma}$. Figure 5 depicts the variation of $q(z)$ versus $z$. From Figure 5, we see that as $z\to-1$, $q\to-1$, while for $z\to\infty$, $q\to(k_{3}-1)$, which may be a positive or negative value according to the value of $k_{3}$. Thus, the derived model can explain both scenarios of the expanding universe, like the early decelerating and late-time accelerating phases. We have estimated the present value of $q(z)$ as $q_{0}=-0.3296$ along CC data and $q_{0}=-0.4195$ for CC+Pantheon data, which reveals the current accelerating phase of the universe. Figure 5 illustrates a phenomenon known as signature-flipping in the evolution of $q(z)$, which is referred to as the transition point $z_{t}$. This transition redshift $z_{t}$ is derived from $q(z_{t})=0$, as
\begin{equation}\label{zt}
	z_{t}=\left[\frac{1}{k_{3}-1}\right]^{\frac{1}{2k_{3}}}-1
\end{equation}
Recently, several cosmologists have reported the value of this transition redshift $z_{t}$ as $z_{t}=0.8596_{-0.2722}^{+0.2886}$ along the SNe Ia dataset and $z_{t}=0.6320_{-0.1403}^{+0.1605}$ along the Hubble data in \cite{refz1}. One is reported as $z_{t}=0.643_{-0.030}^{+0.034}$ in \cite{refz2}, while \cite{refz3} is obtained as $z_{t}=0.646_{-0.158}^{+0.020}$ and $z_{t}=0.702_{-0.044}^{+0.094}$ in \cite{refz4}. In another work, it is reported as $z_{t}=0.684_{-0.092}^{+0.136}$ \cite{refz5,refz6}. We have measured the transition redshift $z_{t}=0.4944$ for CC data, while along CC+Pantheon data, it is estimated as $z_{t}=1.196$. Thus, in comparison to the above results, our measured value of transition redshift $z_{t}$ is consistent with observational datasets.
\begin{figure}[H]
	\centering
	\includegraphics[width=10cm,height=8cm,angle=0]{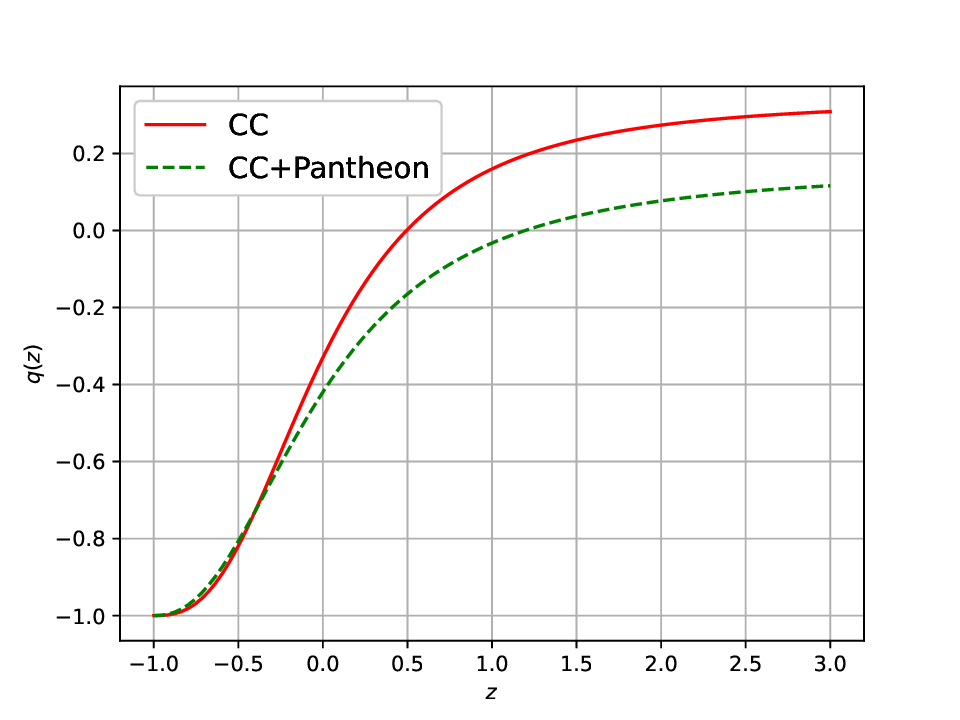}
	\caption{The variation of deceleration parameter $q(z)$ versus $z$.}
\end{figure}
\subsection{$Om(z)$ diagnostic}
The $Om(z)$ diagnostic function classifies cosmic dark energy ideas by behavior. The $Om(z)$ diagnostic function is defined for a spatially homogeneous universe \cite{om}.
\begin{equation}\label{om}
	Om(z)=\frac{\left(\frac{H(z)}{H_{0}}\right)^{2}-1}{(1+z)^{3}-1},
\end{equation}
where $H_{0}$ represents the present value of the Hubble parameter, and $H(z)$ denotes the Hubble parameter as outlined in Eq.\,\eqref{hz}. A positive slope of $Om(z)$ indicates phantom motion, whereas a negative slope signifies quintessence motion. The $\Lambda$CDM model is characterized by the constant $Om(z)$.\\

Equation \eqref{om} delineates the $Om(z)$ function for the model we formulated, while Figure 6 depicts its geometric interpretation.  Figure 6 illustrates that $Om(z)$ ascends with an increase in $z$, signifying a positive gradient.  This indicates that our universe model has characteristics akin to a phantom dark energy hypothesis.  Furthermore, it is evident that when \( z \to -1 \) in the distant future, the value of \( Om(z) \) remains constant, indicating the convergence of our derived model towards the \( \Lambda \)CDM model in the far future.
\begin{figure}[H]
	\centering
	\includegraphics[width=10cm,height=8cm,angle=0]{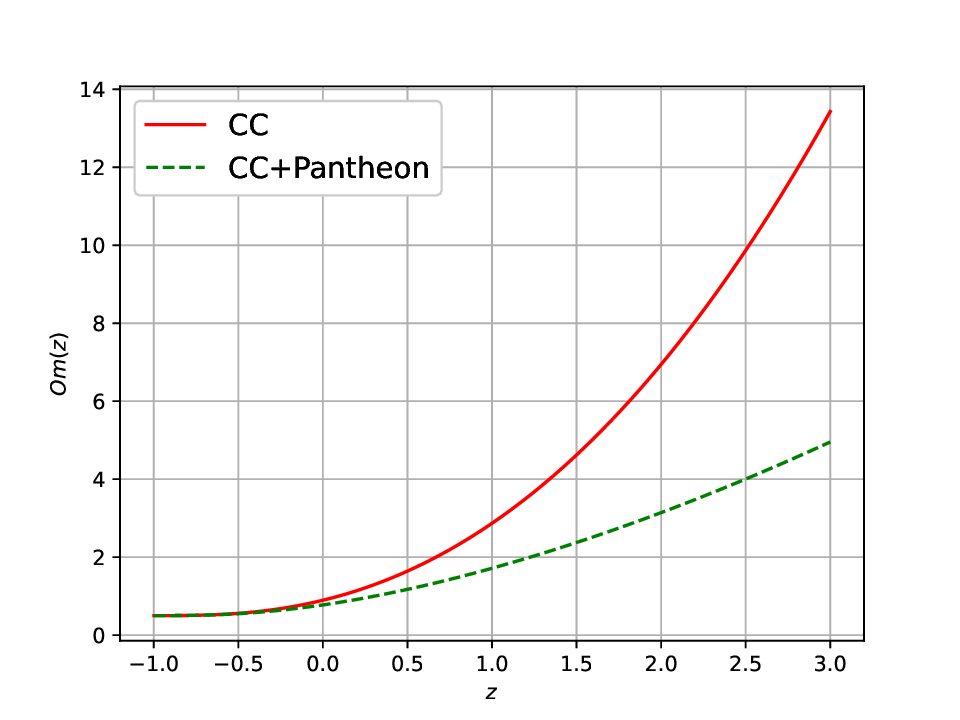}
	\caption{The variation of $Om(z)$ function versus $z$.}
\end{figure}
\subsection{Age of the Universe}
To calculate the age of the universe, we apply the formula below:
\begin{equation}\label{t01}
	t_{0}-t=\int_{0}^{z}\frac{dz'}{(1+z')H(z')}
\end{equation}
By employing the Hubble function expression \eqref{hz} in \eqref{t01}, we derive the age of the universe as
\begin{equation}\label{t02}
	t_{0}=\lim_{z\to\infty}(t_{0}-t)=\lim_{z\to\infty}\left[\frac{\sqrt{2}}{H_{0}}\int_{0}^{z}{\frac{dz'}{(1+z')\sqrt{1+(1+z')^{2k_{3}}}}}\right]
\end{equation}
As \( z \to \infty \), \( t \to 0 \), and \( (t_{0} - t) \to t_{0} \), this represents the current age of the universe. The current estimation of the universe's age is $t_{0}=13.92\pm1.3$ Gyrs based on CC data, whereas for CC+Pantheon data, it is $t_{0}=13.87\pm1.02$ Gyrs. Recent observations indicate that the current age of the universe is reported as $t_{0}=13.7\pm0.02$ Gyr in \cite{spergel2003}, while another source provides an age of $t_{0}=13.772\pm0.059$ Gyr in \cite{age1}.  Consequently, our calculated value for the current age aligns closely with observational findings.

\section{Conclusions}
The current transit universe model constitutes an exact solution to the field equations of a recently introduced Metric-affine $f(R,T,Q,T_m)$ gravity theory [Herko et al. \textit{Phys. Dark Univ.} \textbf{34} (2021) 100886]. We formulate a hyperbolic solution and derive the observational constraints on the model parameters utilizing the most recent observational datasets.  We examine the evolution of cosmological parameters, including the deceleration parameter $q(z)$, the effective equation of state parameter $\omega_{\rm eff}$, the dark energy equation of state parameter $\omega_{\rm de}$, to elucidate the characteristics of the observable universe.  We do the $Om$ diagnostic test for the model, which encapsulates the quintom-A-type scenarios of the model.\\

We have measured the value of the Hubble constant $H_{0}=65.3_{-2.5}^{+2.1}$ km/s/Mpc and the model parameter $\gamma=-0.6510_{-0.011}^{+0.0065}$ for the CC dataset, while for the CC+Pantheon dataset, we measured $H_{0}=69.0\pm1.8$ km/s/Mpc and $\gamma=-0.6648_{-0.0027}^{+0.0022}$. In $f(R)$ gravity theory, this value of the Hubble constant $H_{0}$ is measured as $H_{0}=68.326_{-1.045}^{+1.005}$ km/s/Mpc in \cite{h1}, $H_{0}=69.8\pm1.3$ km/s/Mpc in \cite{h6}, $H_{0}=69.7\pm1.2$ km/s/Mpc in \cite{h7}. In the same gravity theory, \cite{hc1} measured $H_{0}=68.59^{+1.85}_{-1.82}$ km/s/Mpc using the SN+CC dataset, while for the $\Lambda$CDM model, it was obtained as $H_{0}=68.60^{+1.84}_{-1.84}$ km/s/Mpc. The Planck 2018 results \cite{ref7} reported this value as $H_{0}=67.4\pm0.5$ km/s/Mpc. In $f(T)$ gravity theory, the value of the Hubble constant is measured as $H_{0}=68.19^{+1.9}_{-0.93}$ km/s/Mpc in \cite{hc4} using CC+SN, and in $f(Q)$ gravity theory, it is reported as $H_{0}=69.0\pm2.0$ km/s/Mpc in \cite{hc9}. Thus, our estimated values of the Hubble constant are compatible with these measurements. Consequently, our estimated values for the Hubble constant fall within the range of $64< H_{0}<73$ based on the analysis of both datasets.\\

We have used the value of the arbitrary constant $k=1$ in our analysis and discussion of the results. We have obtained the relationships between $\alpha$, $\beta$, and $\gamma$ as $\alpha=\frac{9+13\gamma}{12+18\gamma}$ and $\beta=\frac{9+13\gamma}{6+8\gamma}$. Subsequently, by using the value of $\gamma$, we have calculated the constrained values of the other model parameters $\alpha$ and $\beta$ as $\alpha=0.6780_{-0.1184}^{+0.0583},~0.5246_{-0.0637}^{+0.0277}$ and $\beta=1.9042_{-0.3466}^{+2.7862},~10.6428_{-2.1803}^{+0.3056}$, based on two sets of observational data, CC and CC+Pantheon. Using the above estimations, we found a transit phase quintom-A-type universe model with the current value of the deceleration parameter $q(z)$ as $q_{0}=-0.3296$ along CC data and $q_{0}=-0.4195$ for CC+Pantheon data, which reveals the current accelerating phase of the universe. We have measured the transition redshift $z_{t}=0.4944$ for CC data, while along CC+Pantheon data, it is estimated as $z_{t}=1.196$, which was recently reported as $z_{t}=0.8596_{-0.2722}^{+0.2886}$ along the SNe Ia dataset and $z_{t}=0.6320_{-0.1403}^{+0.1605}$ along the Hubble data in \cite{refz1}. One is reported as $z_{t}=0.643_{-0.030}^{+0.034}$ in \cite{refz2}, while \cite{refz3} is obtained as $z_{t}=0.646_{-0.158}^{+0.020}$ and $z_{t}=0.702_{-0.044}^{+0.094}$ in \cite{refz4}. In another work, it is reported as $z_{t}=0.684_{-0.092}^{+0.136}$ \cite{refz5,refz6}. Thus, in comparison to the above results, our measured value of transition redshift $z_{t}$ is consistent with observational datasets.\\ 

We found that the current value of the effective EoS parameter is $\omega_{\rm eff}=-0.55\pm0.16$ using CC data, and when combining CC with Pantheon data, it is estimated to be $\omega_{\rm eff}=-0.62\pm0.11$, which is similar to the $\Lambda$CDM value of about $\omega_{\rm eff}\approx-0.70$ \cite{ref2,ref3}. We have measured the present value of the dark energy EoS parameter as $\omega_{\rm de}=-1.2648$ along CC data, while for CC+Pantheon datasets $\omega_{\rm de}=-1.6285$, and we found that the evolution of the dark energy EoS parameter $\omega_{\rm de}$ of the model behaves just like a quintom-A universe. In \cite{quintom1}, it was shown that there are two main types of quintom models: quintom-A, where the value of $\omega_{\rm de}$ started from quintessence in the early universe and a phantom field takes over in the late universe, leading to a change from above to below the phantom divide; and quintom-B, where the equation of state moves from below $\omega_{\rm de}=-1$ to above $\omega_{\rm de}=-1$. Thus, one can see that Figure 4 shows quintom-A-like scenarios of the model. The dark energy equation of state $\omega_{\rm de}$ is reported as $\omega_{\rm de}=-0.98\pm0.12$ in \cite{spergel2003}, while it is measured as $\omega_{\rm de}=-1.0\pm0.19$ and $\omega_{\rm de}=-1.10\pm0.14$ in \cite{komatsu2011}. The Om diagnostic test indicates that our derived model is a phantom dark energy model. The current estimation of the universe's age is $t_{0}=13.92\pm1.3$ Gyrs based on CC data, whereas for CC+Pantheon data, it is $t_{0}=13.87\pm1.02$ Gyrs. Recent observations indicate that the current age of the universe is reported as $t_{0}=13.7\pm0.02$ Gyr in \cite{spergel2003}, while another source provides an age of $t_{0}=13.772\pm0.059$ Gyr in \cite{age1}.  Consequently, our calculated value for the current age aligns closely with observational findings.\\

Thus, our derived universe model in the metric-affine $f(R,T,Q,T_m)$ gravity theory shows the quintom-A-like model without inclusion of the cosmological constant, which is not clearly shown in most of the modified theories. This theory is the maximal extension of GR and is based on a generalized metric-affine connection that may succeed in explaining the whole evolution of the universe. Thus, the metric-affine $f(R,T,Q,T_m)$ gravity theory may be helpful in explaining quintom-like universes, which could encourage cosmologists to further investigate it.

\section*{Declaration of competing interest}
The authors declare that they have no known competing financial interests or personal relationships that could have appeared to influence the work reported in this paper.

\section*{Acknowledgments}
We express our gratitude to Inter-University Centre for Astronomy and Astrophysics (IUCAA), Pune, India for providing the resources and assistance necessary to finish this work.

\section*{Data Availability Statement}
No data associated in the manuscript.

\section*{Author Contributions}
All authors contributed to the study conception and design. The first draft of the manuscript was written by [Dinesh Chandra Maurya] and all authors commented on previous versions of the manuscript. All authors read and approved the final manuscript.

	
\end{document}